\documentstyle[12pt,epsf]{article}

\begin{document}
\def\Journal#1#2#3#4{{#1} {\bf #2}, #3 (#4)}
\def\NCA{\em Nuovo Cimento}
\def\NIM{\em Nucl. Instrum. Methods}
\def\NIMA{{\em Nucl. Instrum. Methods} A}
\def\NPB{{\em Nucl. Phys.} B}
\def\PLB{{\em Phys. Lett.}  B}
\def\PRL{\em Phys. Rev. Lett.}
\def\PRD{{\em Phys. Rev.} D}
\def\ZPC{{\em Z. Phys.} C}
\def\st{\scriptstyle}
\def\sst{\scriptscriptstyle}
\def\mco{\multicolumn}
\def\epp{\epsilon^{\prime}}
\def\vep{\varepsilon}
\def\ra{\rightarrow}
\def\ppg{\pi^+\pi^-\gamma}
\def\vp{{\bf p}}
\def\ko{K^0}
\def\kb{\bar{K^0}}
\def\al{\alpha}
\def\ab{\bar{\alpha}}
\def\be{\begin{equation}}
\def\ee{\end{equation}}
\def\bea{\begin{eqnarray}}
\def\eea{\end{eqnarray}}
\def\CPbar{\hbox{{\rm CP}\hskip-1.80em{/}}}

HLRZ 69/96\hfil\\
SHEP 96/26\hfil\\
WUB  96/39\hfil\\\vskip 2em

\begin{center}
   {\bf SPIN AND VELOCITY DEPENDENT CORRECTIONS TO THE INTERQUARK POTENTIAL
AND QUARKONIA SPECTRA FROM LATTICE QCD\footnote{Presented
at the conference on quark confinement and the hadron spectrum
(Confinement 96), Como.}\par}       
   \vskip 2em

{\small\rm GUNNAR S.~BALI}\par
{\small\it Physics Department, The University, Highfield,\\
Southampton SO17 1BJ, England}\par\vspace{1em}
{\small\rm KLAUS SCHILLING, ARMIN WACHTER}\par
{\small\it Forschungszentrum J\"ulich, HLRZ,\\ 52425 J\"ulich,
Germany}\par\vspace{1em}
\relax
\end{center} \par

\begin{center}
{\begin{minipage}{5truein}
                 \footnotesize
                 \parindent=0pt
We present results on
(order $v^2$) QCD interquark potentials from SU(3) lattice
simulations on volumes of up to $32^4$ lattice sites at $\beta=6.2$
and $\beta = 6.0$. Preliminary results on quarkonia spectra
as obtained from these potentials are presented.
\par\end{minipage}}\end{center}
\vskip 2em \par
  
\section{Introduction}
Purely phenomenological or QCD inspired potential models
have been proven to reproduce the observed charmonium ($J/\psi$) and
bottomonium ($\Upsilon$) spectra remarkably well.
The success of such models might be understood from QCD
as a Schr\"odinger-Pauli type Hamiltonian
with spin dependent (sd) and velocity
dependent (vd) 
contributions --- that can be parameterized in terms
of seven independent scalar functions of the quark separation
(potentials) --- has been derived directly
from the QCD Lagrangian~\cite{EFG,gromes,BBMP}.
By computing these potentials nonperturbatively on the lattice,
the set of free parameters within the potential picture can be restricted.
The present investigations correspond to a complete expansion of the
QCD Lagrangian up to order $1/m^2$ in the heavy quark mass $m$ or, 
alternatively, order $v^2$ in the relative velocity between the sources.
The form of the Hamiltonian resembles that of NRQCD~\cite{NRQCD} at
this given order. The only additional assumption that enters is the
instantaneous approximation, i.e.\ the potentials are
assumed to have no explicit dependence on the interaction time.

In our determination of the central potential as well as the
relativistic corrections~\cite{tbp}, we find short ranged Coulomb like
contributions which had not been expected so far in two of the potentials.
The main effect is that the $J/\psi$ ``sees'' a stronger 
effective Coulomb
force than the $\Upsilon$.

\section{The Hamiltonian}
Starting from a Foldy-Wouthuysen transformation of the Euclidean
quark propagator in an external gauge field, the connection between
the static interquark potential $V_0(r)$ and Wilson loops can be
derived. By perturbing the propagator in terms of the
inverse quark masses $m_1^{-1}$ and $m_2^{-1}$ around its static
solution, one arrives at the semi-relativistic Hamiltonian (in the CM
system, i.e.\ ${\mathbf p}={\mathbf p_1}=-{\mathbf p_2}$ and ${\mathbf L}=
{\mathbf L_1}={\mathbf L_2}$),
\begin{equation}
\label{ham}
H=\sum_{i=1}^2\left(m_i+\frac{p^2}{2m_i}-\frac{p^4}{8m_i^3}\right)
+V_0(r)+
V_{\mbox{\scriptsize sd}}(r,{\mathbf L},{\mathbf S_1},{\mathbf S_2})+
V_{\mbox{\scriptsize vd}}(r,{\mathbf p}),
\end{equation}
where the potential consists of a central part and sd and
vd corrections~\cite{EFG,gromes,BBMP}:
\begin{eqnarray}
V_{\mbox{\scriptsize sd}}(r,{\mathbf L},{\mathbf S_1},{\mathbf S_2})
&=&\left(\frac{{\mathbf L}{\mathbf S_1}}{m_1^2}
+ \frac{{\mathbf L}{\mathbf S_2}}{m_2^2}\right)
\frac{V_0'(r)+2V_1'(r)}{2r}\label{sdpo}\\
&+&\frac{{\mathbf L}({\mathbf S_1} + {\mathbf S_2})}{m_1m_2}
\frac{V_2'(r)}{r}
+\frac{S_1^iS_2^j}{m_1m_2}\left(R_{ij}V_3(r)+\frac{\delta_{ij}}{3}V_4(r)\right)
\nonumber
\end{eqnarray}
with $R_{ij}=r_ir_j/r^2-\delta_{ij}/3$ and,
\begin{eqnarray}
V_{\mbox{\scriptsize vd}}(r,{\mathbf p})
&=&\frac{1}{8}\left(\frac{1}{m_1^2}+\frac{1}{m_2^2}\right)
\left(\nabla^2V_0(r)+\nabla^2V_a(r)\right)\\\nonumber
&-&\frac{1}{m_1m_2}\left\{p_i,p_j,S_{ij}\right\}_{W}
+\sum_{k=1}^2\frac{1}{m_k^2}\left\{p_i,p_j,T_{ij}
\right\}_{W}
\end{eqnarray}
with $S_{ij}=\delta_{ij}V_b(r)-R_{ij}V_c(r)$ and
$T_{ij}=\delta_{ij}V_d(r)-R_{ij}V_e(r)$.
The symbol $\{\cdot,\cdot,\cdot\}_{W}=
\frac{1}{4}\{\cdot,\{\cdot,\cdot\}\}$
denotes Weyl ordering of the three arguments.
$V_1'$ -- $V_4$ are related to spin-orbit and spin-spin interactions,
$V_b$--$V_e$ to orbit-orbit interactions and the Darwin-like term
that incorporates $\nabla^2V_a$ modifies the
central potential.
$V_1'$ -- $V_4$ and $\nabla^2V_a$ --
$V_e$ can be computed from lattice correlation functions
of Wilson loop like operators in Euclidean time.
Pairs of the potentials are related by Lorentz invariance
to the central potential
($V'_2-V'_1=V'_0$, $V_b-2V_c=\frac{1}{2}V_0-\frac{r}{6}V'_0$,
$V_c+2V_e(r)=-\frac{r}{2}V_0'$)~\cite{gromes,BBMP},
such that only six out of them are truly independent.

\section{Results}
We find the sd potentials $V'_2$, $V_3$, $V_4$ to be short ranged and
to agree with tree level lattice perturbation theory. The agreement
is even better when a running coupling in momentum space is allowed.
Apart from a constant part, which accounts for the string tension, the
spin-orbit potential $V'_1$ contains an attractive Coulomb-like piece
which is not expected from perturbation theory and makes up about
20~\% of the Coulomb part of the central potential. To leading order
minimal surface approximation, $V_a$ does not contain a long range
component while the short range part is expected to
vanish from weak coupling perturbation theory.
We, however, find it to behave like
$-1/r$ which means that the Coulomb force within $c\bar c$ systems is
enhanced by about 20~\% in comparison to the $b\bar b$ case. $V_b$ -- $V_e$
are found to agree
qualitatively with the combined long and short distance expectations.
By neglecting the very short range behaviour ($r<0.15$~fm), which is
affected by
running coupling effects, the data can be
parameterized as follows:
\begin{eqnarray}
V_0 &=&\kappa r-\frac{e}{r},\quad
\nabla^2V_a=-\frac{h}{r},\quad
V_1'=-\kappa-\frac{d}{r^2},\nonumber\\
V_2'&=&\frac{a}{r^2},\quad
V_3 =\frac{3a}{r^3},\quad
V_4 =8\pi a\delta^3(r),\\\nonumber
V_b&=&\frac{2}{3}\frac{e}{r}-\frac{\kappa r}{9},\quad
V_c=-\frac{e}{2r}-\frac{\kappa r}{6},\quad
V_d=-\frac{\kappa r}{9},\quad
V_e=-\frac{\kappa r}{6}
\end{eqnarray}
with $a=e-d$.
$V_0$, $V_a$, $V_b$ and $V_d$ contain unphysical self energy constants
which have to be subtracted. The parameter values are:
$e=0.321(7)$, $d =0.065(11)$ and $h =3.75(31)\kappa$.

\section{Spectroscopy}
\begin{figure}
\begin{center}
{\epsfysize=9cm\epsfbox{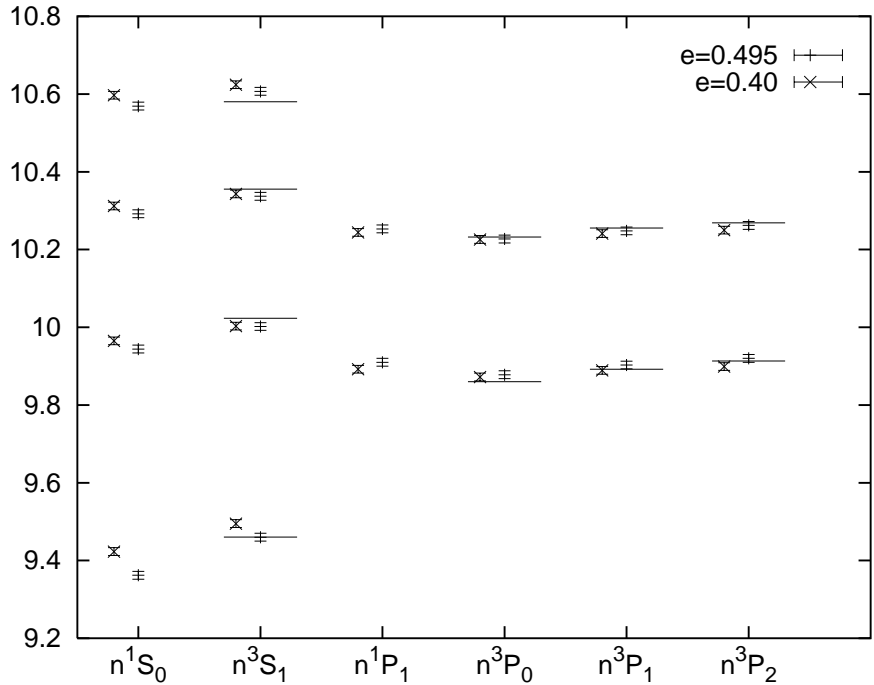}}
\end{center}
\caption{{\small The bottomonium spectrum (in GeV). Horizontal lines are
experimental results.}\label{fig1}}
\end{figure}
Two free QCD parameters have to be determined from experiment, i.e.\
from a fit to the
$\Upsilon$ and $J/\psi$ spectra, namely the heavy quark mass, $m_b$ or
$m_c$, and a scale which sets the initial condition for the running of
the QCD coupling, e.g.\ the parameter $\kappa$.\footnote{
Since the central potential is unbounded, the zero point
energy is in principle arbitrary. However, it turns
out that a zero point energy C in $V_0$ will give rise to counter
terms in $V_b$ and $V_d$ which can be absorbed into a redefinition
of the quark mass~\cite{BBMP}: $m\rightarrow m+C/2$. We
choose $C=0$ by demanding $V_d(0)=0$. Apart from this, the
constituent quark masses of Eq.~\ref{ham} can differ from the
kinetic masses appearing in the dispersion relation to this order in
$1/m$. We find that allowing for such a difference
does not substantially improve the agreement with experiment.}
Since we have not included sea quark effects into
our lattice simulation yet, for the time being, we allow for a free
Coulomb coefficient $e$.
The optimal values of the fit parameters turn out to be
$e=0.495$, $\sqrt{\kappa}\approx 430$~MeV, $m_b\approx 4.79$~GeV
and $m_c\approx 1.41$~GeV
with an error of about $0.1$ on $e$.
Simulations of full QCD~\cite{SESAM} suggest
that values $e\approx 0.40$ are realistic in a world with three
light sea quarks. 
In Fig.~\ref{fig1} the $b\bar b$
spectrum from our best fit ($e=0.495$)
is compared to experiment (horizontal lines) as well as the
result with $e=0.4$.
Fits to improved parametrizations that incorporate the QCD running
coupling are in preparation.
\section*{Acknowledgments}
GSB would like to thank Marshall Baker, Nora Brambilla, Christine
Davies, Lewis Fulcher and Hugh Shanahan for discussions.
During completion of this work GSB has been supported by EU grant ERB
CHBG CT94-0665. We appreciate support by EU grant CHRX CT92-0051 and
DFG grants Schi 257/1-4 and Schi 257/3-2.


\section*{References}
\begin{thebibliography}{99}
\bibitem{EFG}E. Eichten and F. Feinberg, \Journal{\PRD}{23}{2724}{1981}.

\bibitem{gromes}D. Gromes, \Journal{\ZPC}{22}{265}{1984}.

\bibitem{BBMP}A. Barchielli, E. Montaldi,G.M. Prosperi,
    \Journal{\NPB}{296}{625}{1988}; \Journal{\NPB}{303}{752}{1988};
 A. Barchielli, N. Brambilla, G. Prosperi, 
\Journal{\NCA}{103}{59}{1990}.

\bibitem{NRQCD}B.A. Thacker and G.P. Lepage, \Journal{\PRD}{43}{196}{1991}.

\bibitem{tbp}G.S. Bali, K. Schilling, A. Wachter, in preparation.

\bibitem{SESAM}SESAM collaboration: U. Gl\"assner {\em et al.},
\Journal{\PLB}{383}{98}{1996}.

\end{thebibliography}
\end{document}